\documentclass[12pt]{iopart}
\pdfoutput=1
\usepackage{color}
\usepackage{iopams}
\usepackage{graphicx}
\usepackage{subfigure}
\usepackage{cuted}

\usepackage{lineno}

\newcommand{\ds}{\displaystyle}

\begin{document}

\title{Coexistence in neutral theories: interplay of criticality and mild local preferences}
 \author{Claudio Borile}
 \address{DISAT, Politecnico di Torino, Corso Duca degli Abruzzi 24, I-10129 Torino, Italy}
 \author{Daniel Molina-Garcia}
 \address{ Departamento de
    Electromagnetismo y F{\'\i}sica de la Materia e Instituto Carlos I
    de F{\'\i}sica Te\'orica y Computacional. Universidad de Granada.
    E-18071, Granada, Spain}
 \author{Amos Maritan}
 \address{Dipartimento di Fisica `G. Galilei', Universit\`a di Padova, CNISM \& INFN, Via Marzolo 8, 35131 Padova, Italy}
 \author{Miguel A. Mu\~noz}
 \address{ Departamento de
    Electromagnetismo y F{\'\i}sica de la Materia e Instituto Carlos I
    de F{\'\i}sica Te\'orica y Computacional. Universidad de Granada.
    E-18071, Granada, Spain}
\begin{abstract}
  Neutral theories have played a crucial and revolutionary role in fields such as population genetics and biogeography. These theories are critical by definition, in the sense that the overall growth rate of each single allele/species/type vanishes. Thus each species in a neutral model sits at the edge between invasion and extinction, allowing for the coexistence of symmetric/neutral types. However, in finite systems, mono-dominated states are ineludibly reached in relatively short times owing to demographic fluctuations, thus leaving us with an unsatisfactory framework to rationalize empirically-observed long-term coexistence. Here, we scrutinize the effect of heterogeneity in quasi-neutral theories, in which there can be a local mild preference for some of the competing species at some sites, even if the overall species symmetry is maintained.  As we show here, mild biases at a small fraction of locations suffice to induce overall robust and durable species coexistence, even in regions arbitrarily far apart from the biased locations.  This result stems from the long-range nature of the underlying critical bulk dynamics and has a number of implications, for example, in conservation ecology as it suggests that constructing local specific ``sanctuaries'' for different competing species can result in global enhancement of biodiversity, even in regions arbitrarily distant from the protected refuges.
 \end{abstract}

\maketitle

\section{Introduction}
Statistical mechanics models on lattices or networks constitute a deep-rooted theoretical framework in areas of science where the subjects of study are ensembles of many interacting ``building blocks'' such as particles, spins, individuals, agents, and so forth \cite{LiggettInteracting,Gardiner, Krapivsky}.  In particular, the study of genuinely non-equilibrium models, with different types of collective ordering, paved the way for the development of interdisciplinary applications --far beyond tradicional physics problems-- in biology, ecology, epidemiology, and social sciences \cite{Blythe07, Durrett94, Castellano09b, Bikas}.  Within this framework, neutral theories came out --first in population genetics \cite{Kimura,Crow} and then in ecology \cite{Hubbell,Volkov03,Azaele06,Alonso06}, and epidemiology \cite{Pinto}-- as analytically tractable null models, aimed at capturing the main collective and emerging properties of communities of interacting individuals belonging to a limited set of interchangeable types (alleles, species, pathogens, opinions, etc.).  For instance, in the case of population genetics, neutral theories are able to reproduce with remarkable accuracy patterns of relative abundance of different alleles as a result of pure stochasticity and, thus, without making any reference to specific intrinsic differences between them nor to natural selection, leading to a deep conceptual revolution in the field \cite{Kimura}. A similar revolution shattered theoretical ecology after Hubbell's neutral theory of biogeography and biodiversity \cite{Hubbell}.

Different models fall under the common name of ``neutral'' theories; for instance, some of them are spatially explicit, while others are not. However, they all necessarily share two common important traits: symmetry upon species exchange and the existence of absorbing or quiescent states, which account for the constraint that once all individual elements are identical (e.g. a given allele fixated through a population or a mono-dominated forest) the system remains indefinitely unaltered, at least in the absence of mutation, immigration, or other external perturbations.  The most paradigmatic example of this class of models is the exactly solvable \emph{voter model} \cite{LiggettInteracting}, a two-species parameter-free competition model characterized by two symmetric absorbing states (representing the extinction of one species and the subsequent mono-dominance of the remaining one), with very irregular domain frontiers (which in physical terms stem from the absence of surface tension \cite{LiggettInteracting,Dornic01, AlHammal05,Dallasta2007}), and logarithmic coarsening \cite{Krapivsky,Dallasta2008}. It is noteworthy that owing to the symmetry between the two species, the average growth rate of each of them necessarily vanishes, and thus the voter model sits by construction right at a critical point, with diverging characteristic length and time scales \cite{LiggettInteracting,Dornic01, AlHammal05,Canet05}.  Variants of the voter model in which ordered and disordered phases emerge as a control parameter is varied have been studied in the literature (see e.g. \cite{AlHammal05,Castellano09a} and references therein); right at their critical point these models behave like the pure voter model, confirming that this constitutes a robust, {\it generalized voter} (GV), universality class.
 Owing to the existence of strong fluctuations in the dynamics, any finite-size neutral system is doomed to eventually fall into one of the absorbing states.  In the particular case of the voter model the typical time needed to reach absorption, $T$, can be exactly computed. It shows a power-law dependence on the size of the system $T \sim N^\alpha$ with a dimension-dependent exponent $\alpha=\alpha(d)$, with logarithmic corrections at the upper critical dimension $d=2$: $T \sim N \log(N)$ \cite{Frachebourg96, Krapivsky}.  This behavior is shared by models in the GV class at criticality.  Therefore, coexistence in the voter model --in the absence of mutation or immigration-- is just transitory or ``fragile''.  On the other hand, a strong signature of the existence of a phase of robust coexistence would be provided by the observation of exponential scaling of $T$ with $N$ as would correspond to the Arrhenius law for the escape from a potential well \cite{Gardiner}.  If one is interested in the ecological/biological interpretation of neutral theories, the transitory nature of coexistence leaves unanswered the question of how diversity (i.e. alleles/species coexistence) can be preserved over large time scales; thus, one needs to resort to relative large mutation and/or migration rates --which might be unrealistic-- to justify the empirically encountered rich diversity.

 Alternative mechanisms fostering coexistence have been extensively searched-for in the literature. Coexistence can be stabilized by considering the breaking of neutrality at local but not at global scales. For instance: the introduction of negative density dependence in the ability of a species to invade a new territory \cite{Volkov2005,Borile13} or considering quenched environmental conditions which favor each of the species in some regions, but without an overall preference for any of them \cite{Masuda10, Pigolotti10, Borile13}, lead to much larger extinction times ($T \sim \exp(c N)$, where $c > 0$ is a constant) than those of the pure voter model, entailing truly stable or ``robust'' species coexistence.  Similarly, models have been studied where the presence of ``zealots'' --i.e. sites which do not alter their state under any circumstances, thus breaking the local symmetry in a ``hard'' way-- prevents the corresponding absorbing state from be reached, precluding extinction \cite{Mobilia03,Mobilia05,Mobilia07}.  Keeping in mind the ecological interpretation of neutral theories, our aim here is to investigate the effect of locally breaking the neutral dynamics in a ``soft'' way.  More precisely, we introduce a slight local bias towards one of the two states only at a few specific locations, with conflicting local preferences
 existing across the system. Contrarily to the case of the ``hard'' constraint imposed by zealots (which do not ever alter their state), here all sites are allowed to take any of the two states even if there are local biases.

\section{Definition of the model and Mean Field analysis}
For sake of clearness but without lack of generality, we consider the voter model \cite{LiggettInteracting} as a minimal model of neutral competition of two species (generalizations to $S$ species are straightforward \cite{Borile12}). Sites on an arbitrary $d$-dimensional lattice are endorsed with binary variables, $\sigma_i\in\{-1, 1\},\ i\in\mathbb{Z}^d$, encoding the type of species at each location; each node changes its state with a probability proportional to the number of neighbors in the opposite state (see below). Trivially, the model has two symmetric absorbing or mono-dominated states.  The perturbation we consider is in the form of spatial environmental heterogeneities --which constitute a key and unavoidable aspect of real ecosystems \cite{Tilman94}-- that preserve the overall neutral symmetry, as well as the existence of the absorbing states but that locally favor one of the competing species. This is modeled by a quenched external-field ($\tau$) or intrinsic preference for one of the two states at a limited fraction of the sites as follows.  We partition the lattice in three disjoint sets: $\Lambda_+$ where states (``spins'') intrinsically tend to conform with the ``up'' state (i.e. $\tau=+1$), $\Lambda_-$ with an intrinsic preference for the ``down'' state (i.e. $\tau=-1$), and a neutral set $\Lambda_\emptyset$ with no preference ($\tau=0$).
 The relative size, $|\bullet|$, of these three sets is fixed via a parameter $\eta\in[0,1/2]$, \emph{viz.} $|\Lambda_+|=|\Lambda_-|=\eta N$ and $|\Lambda_\emptyset|=(1-2\eta)N$; in particular, the fraction $\eta$ may diminish with system size if a non-extensive amount of biased sites is considered. In continuous time, the model is defined by the flipping rates at any given site $i$: \begin{equation} \begin{array}{lcl}
 W_{\sigma_i\to -\sigma_i}
=\displaystyle \frac{1-\epsilon\tau_i\sigma_i}{2|z(i)|}\sum_{j\in z(i)}(1-\sigma_i\sigma_j),
\end{array}
\label{eq:model}
 \end{equation}
 where $z(i)$ is the set of nearest neighbors of vertex $i$, the ``external fields'' $\tau_i$ are quenched variables taking values $0,+1,-1$ if $i\in\Lambda_\emptyset,\Lambda_+,\Lambda_-$, respectively, and $0 \le \epsilon \le1$ is a constant parameter defining the strength of the local bias. Eq.(\ref{eq:model}) is the sum of a linear term representing the voter model dynamics and a term that lowers or enhances the flipping rate by a constant amount $\epsilon$ depending on whether the change results in alignement with $\tau_i$ or not. For $\epsilon=0$ or for $\eta=0$, we recover the standard voter model. On the other extreme, for $\epsilon=1$, sites with a non-zero external field, $\tau_i$, are always aligned with the field (i.e. are ``zealots''; in this case, the mono-dominated (absorbing) states are explicitly removed, leading to a different family of models \cite{Mobilia03,Mobilia07}). Observe that the model is symmetric in the sense that if the labels of all individuals and the direction of all external fields are switched the system remains unchanged.

 To obtain analytical insight, we first consider the mean-field (MF) version of the model.  For this, we consider the dynamics on a complete --fully connected-- graph where all sites are nearest neighbors of each other, which can be interpreted as a model of mutually interconnected communities (metacommunities) \cite{Metapopulation}. For example, we could think of two competing species occupying an area composed of patches or islands, and individuals can disperse from one to the other; two of  islands could be more favorable for each of the two species respectively, while a third one could be neutral.  In this three-island case the state of the system can be completely defined by three macroscopic variables. Let $x$ be the fraction of sites in the whole system which satisfy the local preference for the up state, i.e. $\sigma_i=\tau_i=+1$, $y$ the fraction of sites satisfying the opposite preference $\sigma_i=\tau_i=-1$, and $z$ the fraction of sites that are in the up state in neutral sites, $\sigma_i=+1, \tau_i=0$. By construction, $x$ and $y$ are defined in $[0,\eta]$ while $z$ is defined in $[0, 1-2\eta]$. In the infinite size limit, the model can be easily verified to be ruled by the set of deterministic equations \begin{eqnarray}
  &&\dot{x}=(1+\epsilon)(\eta-x)(\eta+x-y+z)-(1-\epsilon )x(1-\eta-x+y-z) \nonumber \\
  &&\dot{y}=(1+\epsilon)(\eta-y)(1-\eta-x+y-z)-(1-\epsilon )y(\eta+x-y+z) \nonumber \\
  &&\dot{z}=(1-2\eta -z)(\eta+x-y+z)-z(1-\eta-x+y-z).
\label{eq:deterministic}
\end{eqnarray}

The system above has three fixed points, a standard linear stability analysis reveals that two of them --corresponding to the symmetric absorbing states at $(x, y, z)=(0, \eta, 0), (\eta, 0, 1-2\eta)$-- are unstable while the third one at $(x_*, y_*,z_*)=((1+\epsilon)\eta/2, (1+\epsilon)\eta/2, 1/2-\eta)$ is a stable attractor of the dynamics. The last point corresponds to a state of symmetrical coexistence, implying a non-trivial and rich biodiversity over all the ecosystem, independently of the local bias strength and on the fraction of biased nodes (observe that the limit $\eta \rightarrow 0$ of Eq.(2) is singular, as for $\eta=0$ variables $x$ and $y$ cannot be defined).  This conclusion holds also for large but finite values of $N$ (as can be seen by writing down a Fokker-Plank equation from the microscopic dynamics employing a large-$N$ expansion \cite{Gardiner}) where one obtains a stochastic equation with the same deterministic part plus a sub-leading noise term, confirming that the presence of some non-neutral patches prompts robust coexistence in metacommunities.

\begin{figure}[htp]
 \centering
 \includegraphics[width=0.8\columnwidth]{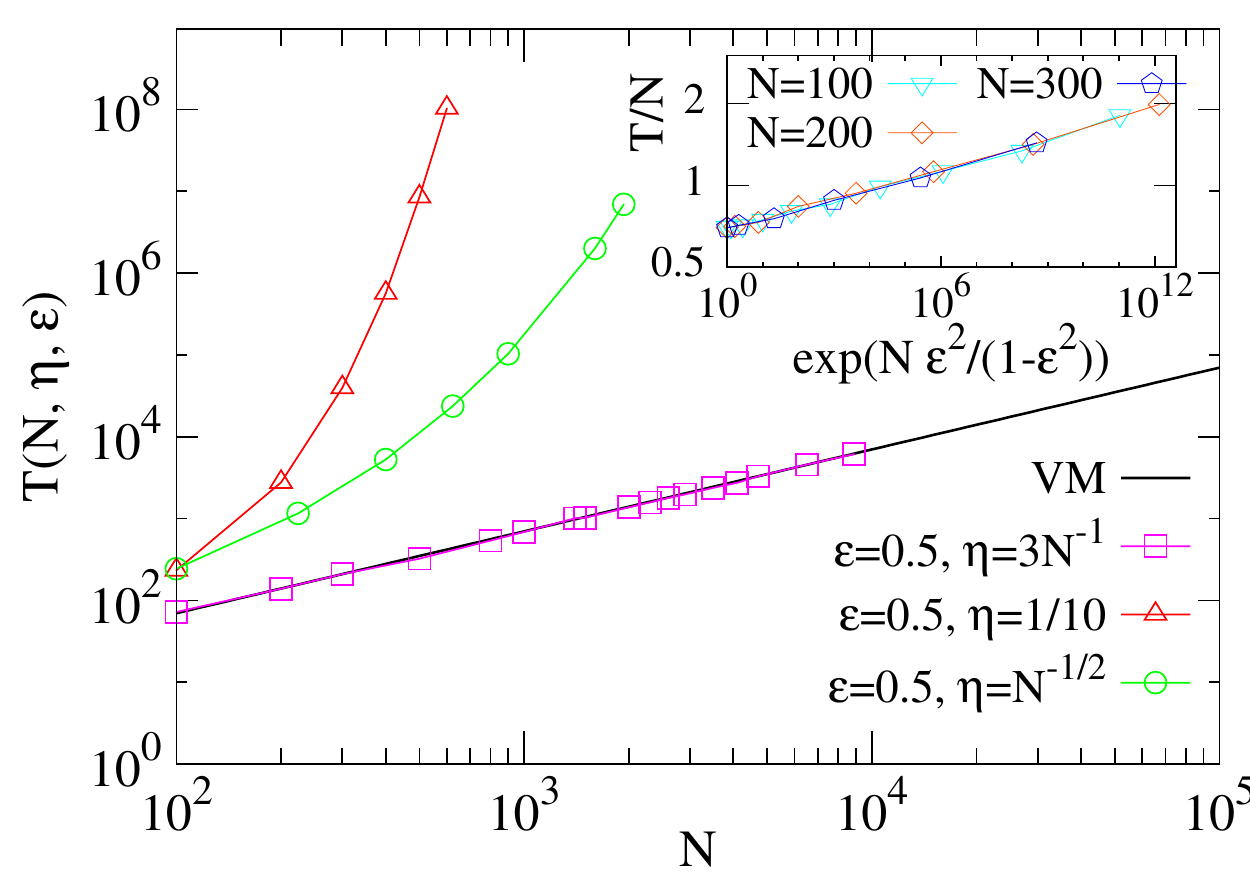} \caption{Mean time to reach the mono-dominated (absorbing or quiescent) state, $T$, in a fully connected (i.e. mean-field) network as a function of system size $N$, for a given strength of the intrinsic preference $\epsilon$, and for extensive and non-extensive values of $\eta N$ (in particular, $\eta=const.=1/10$, $\eta=1/\sqrt{N}$, and $\eta =3/N$). For the non-extensive case, $\eta \propto 1/N$, the perturbation has no effect (in the infinite-size limit) and the coexistence is transitory or fragile, as it is in the pure voter model (VM); $T(N) \sim N \ln(N)$.  Instead, the exponential behavior in the extensive (with $\eta=1/10$) and sub-extensive ($\eta=1/\sqrt{N}$) cases reveals the existence of robust coexistence.  In the inset, we show an example of the collapse of curves obtained varying $\epsilon$ for different sizes of the system in the extensive case using the ansatz in Eq.(\ref{eq:MTAmf}). Averages are performed over, at least, $1000$ realizations.} \label{fig:Fig1} \end{figure}
 \begin{figure}[htp]
  \centering
 \includegraphics[width=0.98\textwidth,height=0.7\textwidth]{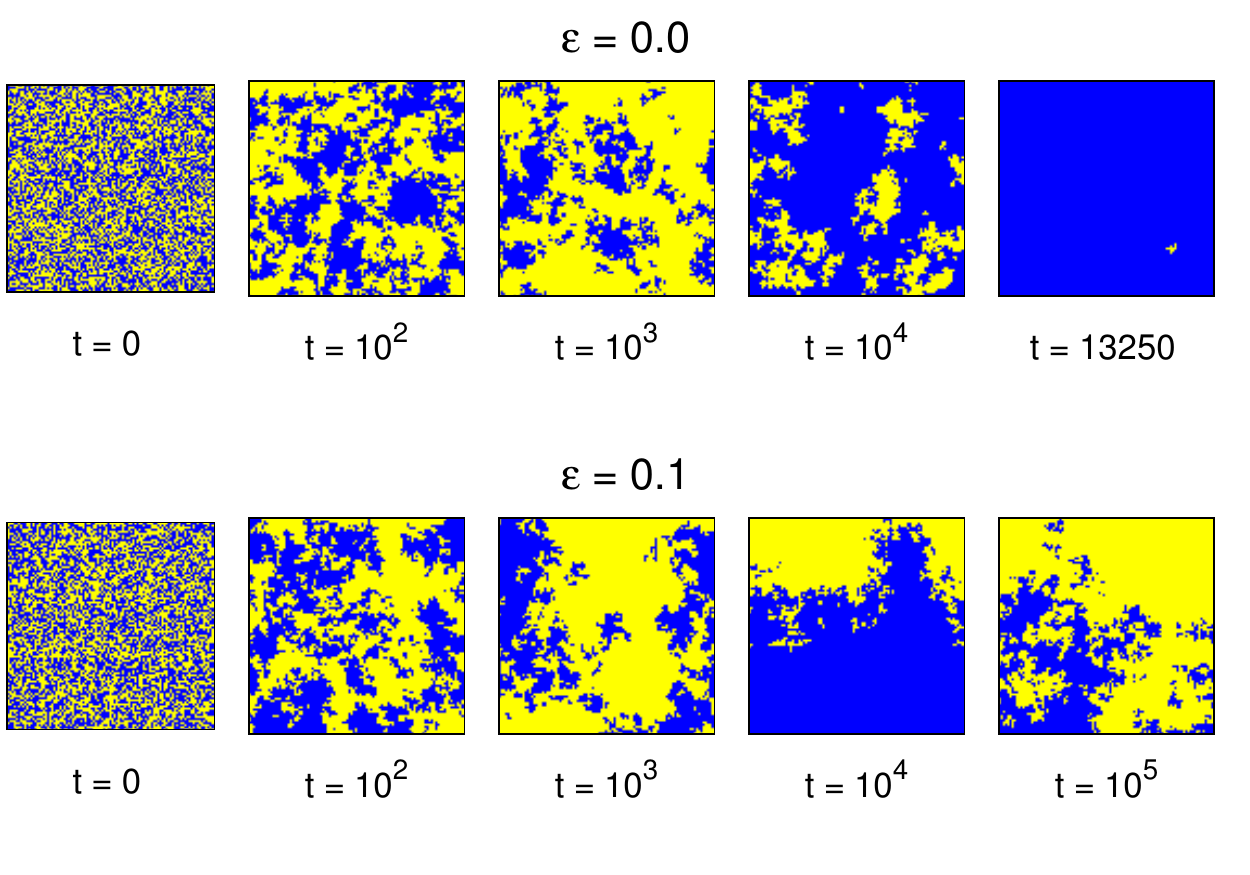}
 \caption{Snapshots of the model evolution at $5$ different times in a two-dimensional square lattice lattice of size $N=L\times L,\ L=50$, with random initial conditions (with periodic boundaries at the right and left borders) and two different values of the strength of the local field acting only at the upper ($\tau=+1$) and lower ($\tau=-1$) boundaries: (top) results for the pure voter model, i.e. $\epsilon=0$ and (bottom) results for boundary field strength $\epsilon=0.1$ (bottom).  Observe, for instance, how in the lower row dark ($-1$) states tend to dominate in the lower half and clear ($+1$) states dominate above. Thus, a mild bias acting only at the system boundaries fosters overall phase coexistence for extremely long times by locally favoring one of the two, otherwise symmetric, species. See Fig.3 for a more quantitative analysis.}
 \label{fig:Fig2}
\end{figure}  
In Fig.\ref{fig:Fig1} we present results of computer simulations using the Gillespie algorithm for the scaling of $T$ as a function of the total number of nodes $N$ on the complete (i.e. fully connected) network for various values of $\eta$.  In the case in which external fields are applied to an extensive number of sites (i.e. $\eta=const.$) it shows a clear exponential scaling, while the non-extensive  case (e.g. keeping a fixed number of biased sites as $N$ is increased, $\eta \propto 1/N$) it is linear as expected for the mean field pure voter model.  A non-trivial situation arises when $\eta N$ is sub-extensive, i.e.  $\eta$ is not a constant nor it decreases as $1/N$, as for instance, $\eta=N^{-1/2}$. As we shall study in detail later, this is for example the case when an external field acts only at some system boundaries in a two-dimensional system. As shown in Fig.\ref{fig:Fig1} the sub-extensive case still shows an exponential scaling of $T$ with $N$, but weaker than the extensive case. Mathematically this is related to the aforementioned singular limit, $\eta \rightarrow 0$.

 To give an estimate of the scaling form of $T$, we can revisit the simpler case in which all nodes in a complete graph are exposed to a bias (a half positive and a half negative, i.e. $\eta=1/2$) that was analyzed by some of us in \cite{Borile13}. In that two-variable case it can be analytically shown that $T$ depends on $N$ and $\epsilon$ via the approximate relation $\ln T\sim N\epsilon^2/(1-\epsilon^2)$ plus sub-leading corrections \cite{Borile13}. In the present case, we need to include the presence of only a limited number of biased sites, as encoded in the parameter $\eta$. Taking logs on both sides, the simplest ansatz one can employ to study the generic situations 
with any value of $\eta$ is 
\begin{equation}
\ln T\sim \ds  \eta^\alpha N\frac{\epsilon^2}{1-\epsilon^2} + \ln  T_{pure}(N)
\label{eq:MTAmf}
\end{equation}      
where we have replaced heuristically $N$ by a reduced effective size, described by $\eta^\alpha N$, where
$\alpha$ is an unspecified positive constant and the sub-leading correction $T_{pure}(N)$ gives the 
value of $T$ in the pure version of the model (i.e. for $\epsilon=0$). 
Writing $\eta=kN^{\alpha'}$ ($\alpha' \leq 0$) and removing sub-leading term $\ln T_{pure}(N)$ 
we can rewrite the equation above in a more compact form 
\begin{equation}
\ln T\sim \ds N^\zeta\epsilon^2/(1-\epsilon^2)$ where $\zeta=1 + \alpha \alpha'<1,
\label{eq:MTAmf2}
\end{equation}      
which leads to a quite good curve collapse for fully connected (i.e. mean-field) networks (see the inset of Fig.1).

\section{Biased boundaries in two dimensions}
To go ahead and study the consequences of mild biases on spatially-explicit systems, going beyond mean-field predictions, we have considered the model of Eq.(\ref{eq:model}) in a two-dimensional square lattice where the biased sets $\Lambda_+$ and $\Lambda_-$ are taken to be two one-dimensional chains with $L=\sqrt{N}$ sites each, located at the upper and lower boundaries respectively. This corresponds to the sub-extensive case studied above, with $\eta=1/\sqrt{N}$.  Periodic boundary conditions are assumed along the other direction. Fig.\ref{fig:Fig2} portrays an example of a single realization of the stochastic process with and without the soft boundary biases. In the bulk, the dynamics of the system is identical in both cases but --owing to the boundary effects-- the biased system reaches the absorbing state in a much longer time and effectively stays in an active/coexistence state. Indeed, as shown in Fig.\ref{fig:Fig3}, robust, exponential coexistence can be observed, obeying the general collapse formula as that of Eq.(\ref{eq:MTAmf2}), with
 \begin{equation} \ds \ln T^{(d=2)}\sim N^{3/4}\frac{\epsilon^2}{1-\epsilon^2} + \log( N\ln N ) 
\label{eq:expscale} 
\end{equation}
where the last term is simply $\ln T_{pure}(N)$ for the two-dimensional standard voter model \cite{LiggettInteracting,Krapivsky}.  To further investigate the origin of the non-trivial $N^{3/4}$ factor we performed simulations on a rectangular system, where $N=L_{\bot}\times L_{\|}$ to discriminate the effective role of the two directions ($ L_{\|}$ is the length of the biased boundaries and $L_{\bot}$ the distance between them). Results are shown in Fig.\ref{fig:Fig4} and show that a good collapse --even if not of the same quality as above-- is observed replacing $N^ {3/4}$ by $L_{\bot}^{1/2} L_{\|}$ (note that for a square lattice: $L_{\bot}=L_{\|}=\sqrt{N}$, $L_{\bot}^{1/2} L_{\|}= N^{3/4}$). This suggests that characteristic times grow linearly with the size of the biased walls and proportionally to the square-root of the distance between them.  While the linear dependence on $ L_{\|}$ seems intuitive, thus far we have not been able to explain the square-root dependence on $L_{\bot}$.

\begin{figure}[htp]
 \centering
 \includegraphics[width=0.9\columnwidth]{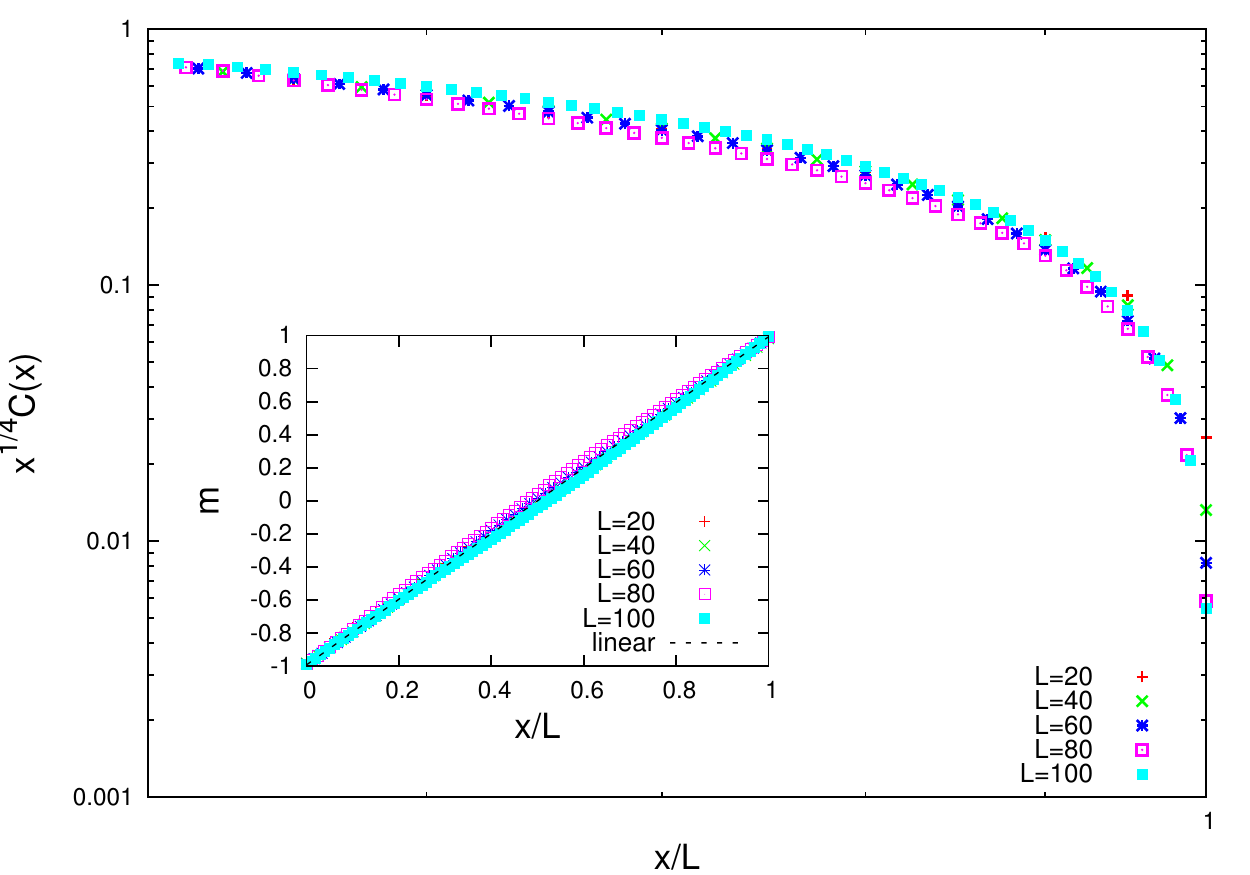}
 \caption{Power-law scaling collapse of the two-point correlation function 
   and mean magnetization along the non-symmetric axis on a two-dimensional lattice (of linear size $L$) with biased boundaries, as defined in equation (\ref{eq:correlations}), both of them plotted as a function of the re-scaled distance $x$ from the wall with negative bias.
   In the bulk the linearity of the magnetization is not perfect due to the highly fluctuating dynamics of the voter model.  In both plots, scaling are not perfect owing to the relatively small system sizes reported and thus to the persistence of corrections to scaling.}
\label{fig:Fig2b}
\end{figure}

\begin{figure}[htp]
 \centering
 \includegraphics[width=0.8\columnwidth]{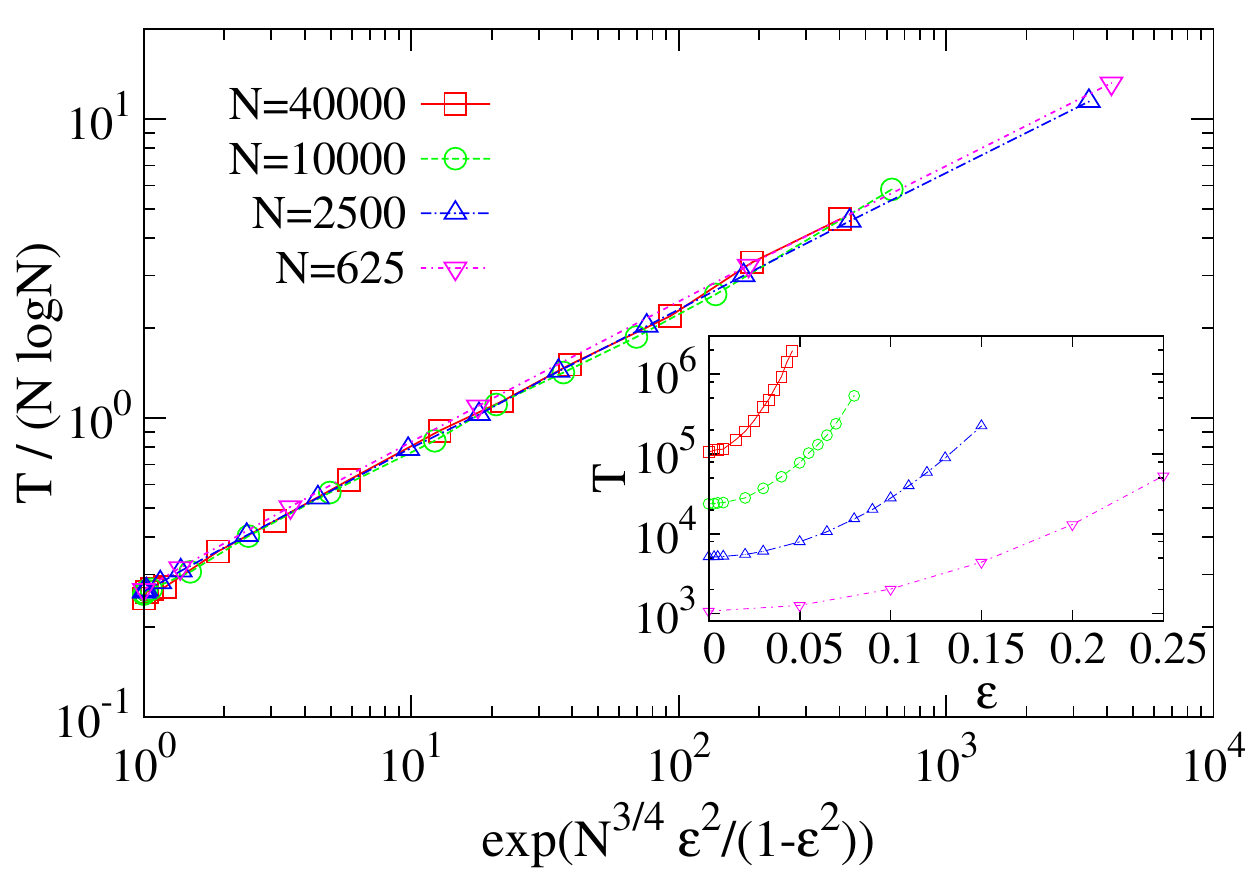}
 \caption{Collapse of the mean time to reach the absorbing state based on the exponential scaling ansatz in Eq.(\ref{eq:expscale}) for various values of $\epsilon$ at fixed $N$. Different curves are for different values of $N$ from $N=10^2$ to $N=10^4$. In the inset the same results are shown without re-scaling. Averages are performed over at least $1000$  realizations.}
\label{fig:Fig3}
\end{figure}

\begin{figure}[htp]
 \centering
 \includegraphics[width=0.8\columnwidth]{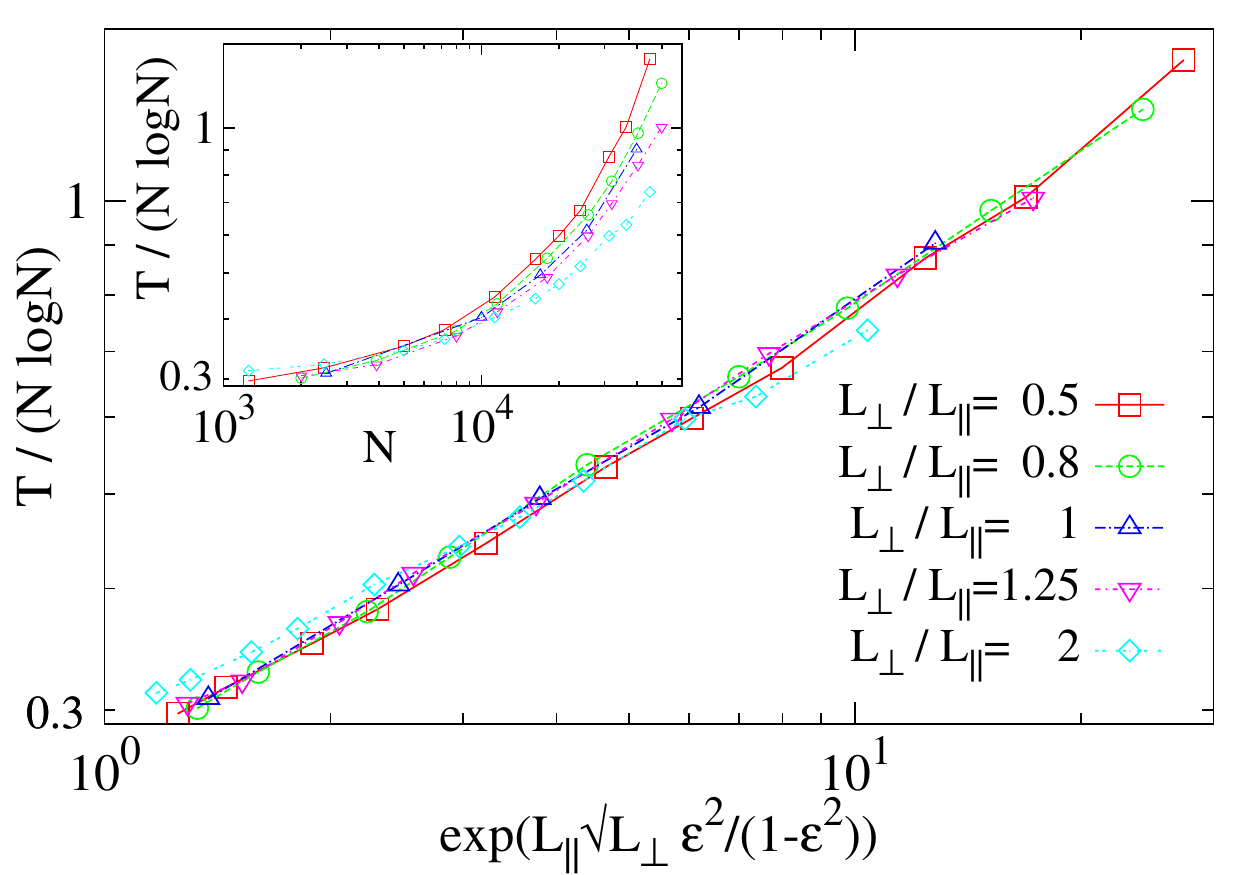}
 \caption{ As in Fig.(\ref{fig:Fig3}) but for rectangular landscapes with opposed biased boundaries of length $L_\|$, separated by a distance $L_\bot$ (note that $\epsilon$ is kept fixed constant, $\epsilon=0.03$, while in Fig.(\ref{fig:Fig3}) it was variable). Averages are performed over at least $1000$ realizations.} \label{fig:Fig4} \end{figure}

 Finally, we have also measured the average value of $\sigma$ (which in physical terms is the ``magnetization'', $m$) as a function of $x_\bot\in[0, L_\bot]$, the lattice position along the $\bot$ direction. At the biased boundaries $m$ is very close to $\pm 1$ depending of the respective external field, while in the bulk it varies linearly with the distance to the boundaries, showing that boundaries propagate their (short-range) influence at arbitrarily large distances. We have also measured two-point correlation function and confirmed the presence of power-law, i.e. scale-free, decays with distance. Both the magnetization $m(x_\bot)$ and the correlation function $C(x_\bot)$ are computed averaging over the direction without biased boundaries ($\|$). These quantities are thus defined as
\begin{equation}
\begin{array}{ll}
m(x_\bot)\equiv <\sigma_{x_{\bot}}>_{\|}\\
C(x_\bot)\equiv <\sigma_{x_{\bot}}\sigma_{L_\bot/2}>_{\|}.
\end{array}
\label{eq:correlations}
\end{equation}
Clearly, all these are consequences of the bulk dynamics being critical, i.e. lacking a characteristic correlation length (Figure \ref{fig:Fig2b}).  Therefore, the interplay between mild biased at distant boundaries and bulk criticality affects the whole system and changes its overall properties, inducing, in particular, stable coexistence.   

\section{Discussion}
In summary, we have investigated the robustness of the critical voter-model behavior upon introducing soft biases which break locally the neutral ($\mathbb{Z}_2$) symmetry. We have shown, both at a mean field level and in spatially explicit two-dimensional systems, that, as long as the number of biased sites grows with system size either extensively or sub-extensively, this type of bias promotes the existence of a well-defined active quasi-stationary state, i.e. they stabilize coexistence between the two competing species even in regions arbitrarily far apart from the biased boundaries.  In particular, we have shown that mild biases at some locations can change the dependence of the characteristic extinction times on system size from power-law (with logarithmic corrections in $d=2$) to exponential, thus preventing the collapse towards the mono-dominated state and greatly enhancing the coexistence of competing neutral species.  This long-ranged global effect stems from the critical, i.e. scale-free, nature of the underlying neutral dynamics in the bulk.  Our results are robust to the introduction of non-symmetrical biased, i.e. stronger for one of the species, except for the fact that the state of coexistence is no longer symmetric.

From the theoretical side, the two-dimensional situation discussed above bears some similarities with wetting phenomena \cite{wettingM,albano,non-eq-wetting,santos2003}. In wetting problems boundary effects can control bulk features arbitrarily far from them \cite{wettingM,albano,non-eq-wetting,santos2003}, but, on the contrary to standard wetting problems, here interfacial descriptions are not useful, as well-defined interfaces separating two different phases are not well defined, i.e. they are too rough and with plenty of overhangs. Thus, theoretical descriptions of the phenomena described here remain elusive. In a future work we shall try to shed further light on these problems by analyzing a field-theoretical version of the voter model \cite{AlHammal05,Canet05,Munoz97,Vazquez08} equipped with adequate boundary conditions.

To conclude, let us remark that our findings here have a number of interesting implications in conservation ecology --where the concept of ``distance of edge influence'' quantifying the spatial scale up to which boundaries in fragmented environments have an impact, is highly relevant \cite{Ries}-- as well as in epidemics and social sciences where neutral dynamics plays a relevant role.  In particular, it suggests that constructing local specific ``sanctuaries'' for each of the competing species in a given community can result in global enhancement of biodiversity, even in regions arbitrarily distant from the preserved refuges.

\section*{Acknowledgments}
  We acknowledge J. de Andaluc\'{i}a project of Excellence P09-FQM-4682, the Cariparo foundation and the `A. Gini' Foundation for financial support. We thank C. Castellano and J. R. Banavar for illuminating discussions and J. Hidalgo, P. Villa and P. Moretti for a careful reading of the manuscript. 

\section*{Bibliography}
\bibliographystyle{unsrt}
\bibliography{Bibliography_new}
  
\end{document}